          [2001/04/25 1.1 (PWD)]
\documentclass[letterpaper,twocolumn]{esapub} % American paper
\input epsf.tex
\usepackage{times}
\usepackage{natbib}
\usepackage{graphics}
\usepackage{epsfig, rotating, latexsym}
\title{Origins of the ALMA project in the scientific visions of the North American, European, and Japanese astronomical communities}
\author{Paul A. Vanden Bout}

\affil{{\it National Radio Astronomy Observatory, 520 Edgemont Road, Charlottesville VA 22903 USA, Email: pvandenb@nrao.edu}}

\begin{document}


\maketitle

\begin{abstract}
ALMA is a worldwide project, the synthesis of early visions of astronomers in its three partner communities, Europe, North America, and Japan. The evolution of these concepts and their eventual merger into ALMA are discussed, setting the background for the papers which follow on the scientific requirements and expected performance of ALMA for extra-galactic, galactic, and solar system research.
\end{abstract}

\section{Introduction}

The Atacama Large Millimeter Array (ALMA) is a fusion of ideas, with roots in the conceptual projects of three astronomical communities: the Millimeter Array (MMA) of the United States, the Large Southern Array (LSA) of Europe, and the Large Millimeter Array (LMA) of Japan.  ALMA has a large collecting area of 7238 m$^{2}$ provided by 64 antennas of diameter 12m.  ALMA is very flexible.  Its antennas can be placed in configurations with sizes from 150 m to 14 km, providing a range of angular resolution of nearly a factor of 1000 at fixed observing frequency. It has the potential to cover all ten frequency bands from 30--950 GHz where the earth's atmosphere is reasonably transparent, with an initial set of receivers that covers the four of these bands.  It has a powerful and flexible signal correlator that can process 2016 baselines with 16 GHz of bandwidth per antenna.  In addition, it has been agreed that ALMA include the Atacama Compact Array (ACA), an array of 12 antennas of diameter 7m (with 4 additional 12m diameter antennas for single dish observations and calibration purposes), equipped with the same receivers as the large array, and equipped with its own signal correlator of similar power as the large array, as well as three additonal receiver bands for all 80 antennas.  The ACA provides data on spatial frequencies  between the compact configuration of the large array and a single (12m) antenna.  ALMA with the ACA and additional receiver bands is known as ``Enhanced ALMA''.  These specifications are those necessary to meet scientific requirements.  They also reflect the union of visions for the three earlier conceptual projects now realized in ALMA.

\section{The Roots of ALMA}

\subsection{Millimeter Array}

The origins of the Millimeter Array (MMA) are found in the pioneering science of the NRAO 36-Foot Telescope (later known as the 12-Meter Telescope), soon followed by the 4.9m telescopes at the University of Texas and Aerospace Corporation, the 14m telescope at the Five Colleges Radio Astronomical Observatory, and the 7m telescope at AT\&T Bell Labs.   The millimeter interferometers of the University of California (Berkeley) at the Hat Creek Observatory (later the Berkeley-Maryland-Illinois Association, or BIMA) and the California Institute of Technology at the Owens Valley Radio Observatory demonstrated the power that comes with high angular resolution for studying the sources found with the single dishes. The experience of using a powerful, flexible array that was provided by NRAO's Very Large Array (VLA) at longer wavelengths was also very influential --- the prime characteristic of the MMA was the ability to obtain rapid high-quality images at 230 GHz, that is, the MMA was to be a millimeter version of the VLA.  The science targets of the MMA included the same broad range of topics seen at the VLA: sun, solar system, stars, galactic interstellar medium, external galaxies, and cosmology.     

\begin{sidewaysfigure*}
\centering
\epsfig{file=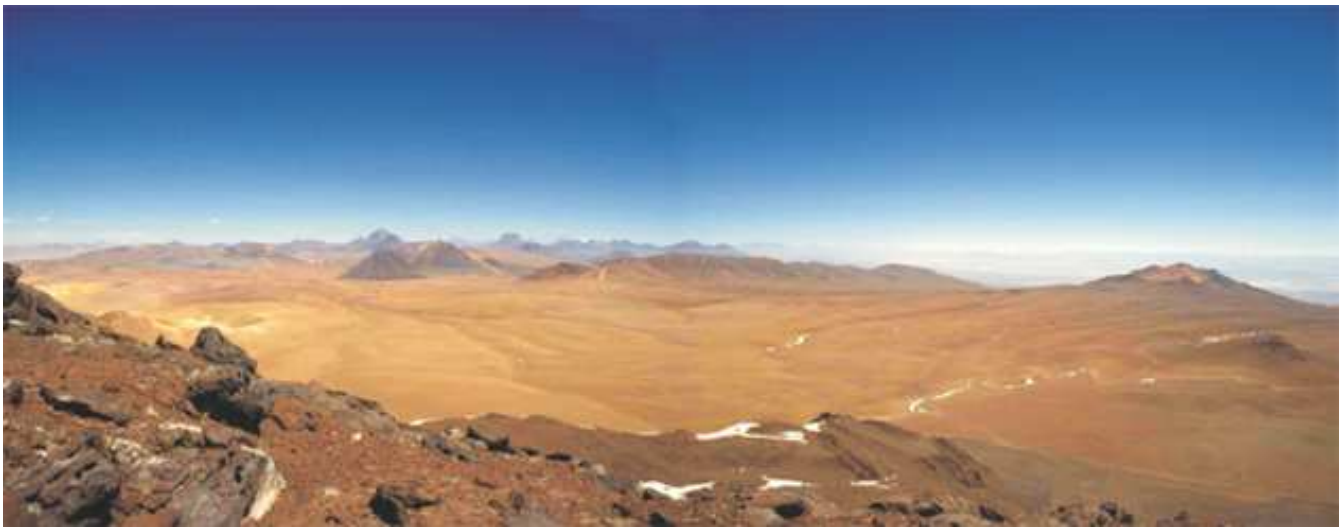,width=7truein} %color figure
\caption{The ALMA site on the Llano de Chajnantor, elevation 5000m, as seen from Cerro Chajnantor (S. J. E. Radford).}
\label{fig}
\end{sidewaysfigure*}

In 1982 an NSF committee appointed to make recommendations for the future of millimeter-wavelength astronomy in the United States called for the development of an interferometer with a collecting area of 1000--2000 m$^{2}$ capable of working at 1mm wavelength with 1$^{\prime\prime}$ resolution at 115 GHz.  The first concept for this interferometer was presented in MMA Memorandum \#1 [1], where an array of 15 antennas of diameter 10m at the VLA site in New Mexico was proposed at an estimated cost of \$36M(US).  The proposal to build the MMA, submitted by Associated Universities, Inc. (AUI) to the National Science Foundation (NSF) in July, 1990, called for an array of 40 antennas of diameter 8m, with four receiver bands covering the atmospheric windows from 30--350 GHz, configurable in four arrays of size 70--3000 m, with a correlator capable of processing 2 GHz per antenna [2].  The proposal discussed two possible sites for the MMA, both in the southwestern United States.  Studies of the atmopheric transparency and phase stability at these sites led to similar studies on Mauna Kea, in Hawaii. Extensive atmospheric monitoring was also conducted there.  Concerns with the limited size of the area available to the MMA on Mauna Kea and with potential environmental problems prompted a search for potential sites in Chile.  The Goddard Institute for Space Studies (later, Harvard-Smithsonian Center for Astrophysics (CfA)) survey of galactic CO emission using 1.2m telescopes, one in each hemisphere, had experienced excellent observing conditions at Cerro Tololo.  In April of 1994 the observatories at Cerro Tololo, ESO La Silla and Paranal, Las Campanas, and  high-elevation sites further inland, including sites previously identified as possibilities for the Sub-Millimeter Array (SMA) of the CfA, were visited.  This search followed by nearly 10 years the first suggestion, in MMA Memorandum \#25 [3] that the MMA be built in the Southern Hemisphere.  The quality of the sites prompted an extension of MMA capability to the submillimeter [4].  The site in Chile selected for the MMA, not one of the sites considered for the SMA, was to the east of the village of San Pedro de Atacama in the Andean altiplano at an elevation of 5000m. This site, named  Llano de Chajnantor, is shown in Fig~\ref{fig}. It was formally proposed to the NSF by AUI as the MMA site in 1996.     The estimated cost of the MMA was \$120M(US). 

\subsection{Large Southern Array}

As in the United States, a broad science program in millimeter wavelength astronomy had developed in Europe, centered around the two telescopes of IRAM, a 30m single dish and an interferometer of three (now six) 15m antennas, the 14m telescope of the Onsala Space Observatory (OSO), and the submillimeter capability of the 15m James Clerk Maxwell Telescope.  The group at the University of Bordeaux were pioneers in millimeter wavelength interferometry.  The first concept for a millimeter interferometer in the Southern Hemisphere came in the late 1980s out of OSO following the success of the Swedish-ESO Submillimetre Telescope (SEST), and called for an array of 10 antennas of diameter 8m to be located near ESO's Very Large Telescope on Cerro Paranal in northern Chile [5]. The estimated cost was \$50M(US).  An array in the Southern Hemisphere became the hallmark of the European array for scientific reasons (the Galactic Center, Magellanic Clouds, etc.) and because ESO, the natural organization for a European astronomical project, had its telescopes there.
 
The discovery of CO emission in a galaxy at a redshift of $z=2.3$, had a profound influence on the size of the LSA. In recognition of the possibility that this galaxy was either atypically luminous or, more likely,  gravitationally lensed, it was argued [6] that a collecting area of at least 10 times that of the IRAM Interferometer on the Plateau de Bure, that is, an unprecedented 10000 m$^{2}$ of collecting area, was required to be able to observe the entire (unlensed) population of such galaxies.  These ideas were incorporated in the thinking for the Large Southern Array (LSA); it's concept proposal (1995) called for a 10000 m$^{2}$ collecting area provided by 50 antennas of diameter 16m or 100 antennas of diameter 11m.  The LSA was to work at frequencies of 350 GHz and below and be equipped with state-of-the-art receivers and signal correlator.  To obtain angular resolution of 0.1$^{\prime\prime}$ at a wavelength of 2.6 mm, configurations of size $\sim$ 10km were contemplated.  Because its highest operating frequency was 350 GHz, the LSA did not require a site as high as that picked for the MMA, and sites at lower elevations of 3300m and 3750m were studied.  The estimated cost of the LSA was about \$250M(US).  

\subsection{Large Millimeter Submillimeter Array}

In Japan, plans for a large millimeter wavelength array grew naturally out of a desire to expand the Nobeyama Millimeter Array.  The Large Millimeter Array (LMA) was discussed in 1983, just following the dedication of the NRO, and in its first form expanded the five 10m diameter antennas of the NRO interferometer to 30, working to a maximum frequency of 230 GHz on baselines up to 1 km.  It was decided in 1987 to expand the concept to 50 antennas of diameter 10 m working at frequencies of 35--500 GHz with the possibility of going to submillimeter frequencies, in configurations of size 20--2000 m [7].  Japanese university groups established a small, automated submillimeter telescope on Mt. Fuji and a small millimeter telescope in Chile.  The site of the NRO precluded observations in the submillimeter, and sites on Mauna Kea and in North Africa and Chile were considered as possibilities. Serious site studies in Chile began in 1992 with a survey of 20 possibilities.  

As the quality of the sites in Chile became apparent, consideration of Mauna Kea was dropped and the prospects of observing in the submillimeter band became the focus of the LMA program, leading to the change in project name to the Large Millimeter/Submillimeter Array (LMSA). In 1995 a memorandum of understanding between the NOAJ and NRAO was signed whereby the two groups agreed to work cooperatively on site studies.  Sites at Pampa la Bola and Rio Frio received intensive study, with Pampa la Bola to the north-east of Llano de Chajnantor site becoming the site of choice in 1997; the importance of 10 km baselines to a combined MMA+LMSA had become clear in a workshop held in Tokyo on submillimeter astronomy at 10 milli-arcseconds resolution.  The Pampa la Bola site showed excellent phase stability and is now the location for a long arm of ALMA antenna pads that stretches northeast of the Llano de Chajnantor.   In 1994 the LMSA received high-level governmental endorsement as a top-priority project for new ground-based astronomical facilities.  Proto-planetary disks and high-z galaxies were considered to be the main scientific targets of the project.  The frequency range of the LMSA was expanded to include 650 and 900 GHz bands in the submilllimeter [8].   

\section{Conclusion}

Few of the participants in the MMA, LSA, and LMSA projects were entirely comfortable with the prospect of building three separate large millimeter/submillimeter arrays in Chile, even if each satisfied to a significant extent the particular wishes of its community. Working groups were established in the General Assemblies of URSI (1993) and the IAU (1994) to study potential  partnerships and discussions on the subject occured frequently.   The major breakthrough occured with the signing of a resolution between ESO and NRAO on June 26, 1997, whereby the two parties agreed to pursue a common project that merged the MMA and LSA into what would eventually be named ALMA. The merged array combined the sensitivity of the LSA with the frequency coverage and superior site of the MMA [9].  The merger was made official in June 1999 with the signing of the Phase 1 ALMA Agreement.  ESO and NRAO worked together in technical, science, and management groups to define and organize a joint project between the two observatories with participation by Canada and Spain.  A flurry of resolutions and agreements ensued, including the choice of ``Atacama Large Millimeter Array'', or ALMA, for the name of the new array in March of 1999.  This effort  culminated in the signing of the ALMA Agreement on February 25, 2003, between the North American and European parties.  

Following mutual discussions over several years, the ALMA Project received a proposal from the NAOJ whereby Japan would provide the ACA and three additional receiver bands for the large array, to form Enhanced ALMA. Further discussions between ALMA and the NAOJ led to the signing of a high-level agreement on September 14, 2004, that makes Japan an official participant in Enhanced ALMA.  Final negotiations on an operations plan for Enhanced ALMA are expected to be concluded by the end of 2005.  ALMA is budgeted at \$552M(Y2000US), shared 50:50 between North America and Europe.  The value assigned to the Japanese contribution to Enhanced ALMA is \$180M(Y2000US).  

In 2004, the last of a number of steps was completed in the long process to secure long-term (50 year) access to the ALMA site with the issuance of three decrees by the Republic of Chile and signing of two contracts. This process included the establishment of AUI in Chile with the same rights and priviledges as ESO, securing by AUI of exploratory mining rights for the site, establishment by Chile of a Science Reserve that includes the site, permission for ESO to open a new observing site,  approval of an environmental impact study, purchase of land for the mid-level operations facility, lease and right-of-way agreements for the site and access road,  approval of a quiet/coordination zone around the site to protect against radio frequency interference, an agreement with the regional government concerning support of cultural and educational activities, and an agreement with CONYCIT concerning the share of observing time on ALMA for Chilean astronomers  and support for the development of astronomy in Chile.  More than 14 government agencies in Chile were involved in the negotiations.

Assuming all three partners are able to meet their commitments, the final project will be cost-shared 37.5\%/37.5\%/25\% between North America, Europe, and Japan, respectively. The observing time, after a 10\% share for Chile, will be shared accordingly.  When Enhanced ALMA, to be known as the Atacama Large Millimeter/Submillimeter Array, with the same acronym ALMA, comes into full operation it will truly be a world millimeter/submillimeter array.

\section*{Acknowledgments}

The author is grateful to  R. S. Booth, R. L. Brown, E. Hardy, M. Ishiguro, and P. A. Shaver for providing information on the early history of the Large Southern Array, Millimeter Array, and Large Millimeter Submillimeter Array.

\section*{References}

%\bibliography{}

\begin{small}

%\begin{thebibliography}{}

1.  Owen, F. N., The Concept of a Millimeter Array, in {\it MMA Memorandum Series}, \#1, 1982.\newline [http://www.alma.nrao.edu/memos/index.html]

%\bibitem{brown94}
2.  Brown, R. L.,  The Millimeter Array, in {\it Astronomy with millimeter and submillimeter wave interferometry}, ed. M. Ishiguro \& W. J. Welch (San Francisco: ASP), ASP Conf. Ser., Vol. 59, 398--404, 1994.

3.  Gordon, M. A., Are we thinking boldly enough?, in {\it MMA Memorandum Series}, \#25, 1982.\newline [http://www.alma.nrao.edu/memos/index.html]

%\bibitem{brown98}
4.  Brown, R. L.,  Technical specification of the Millimeter Array, in {\it Advanced technology MMW, radio, and terahertz telescopes}, ed. T. G. Phillips (Bellingham: SPIE), Proc. SPIE, 3357, 231--7,  1998.

%\bibitem{booth94}
5.  Booth, R. S.,  A Southern Hemisphere Millimetre Array, in {\it Astronomy with millimeter and submillimeter wave interferometry}, ed. M. Ishiguro \& W. J. Welch (San Francisco: ASP), ASP Conf. Ser., Vol. 59, 413--8, 1994.

%\bibitem{downes98}
6.  Downes, D.,  New Directions for Millimeter Astronomy in the 21st Century, in {\it Frontiers of space and ground-based astronomy}, ed. W. Wamsteker, M. S. Longair, \& Y. Kondo (Dordrecht: Kluwer), Astrophysics \& Space Science Library, 187, 133--43,  1994.

%\bibitem{ishiguro94}
7.  Ishiguro, M., Kawabe, R., Nakai, N., Morita, K.-I., Okumura, S. K., and Ohashi, N.,  The Large Millimeter Array, in {\it Astronomy with millimeter and submillimeter wave interferometry}, ed. M. Ishiguro \& W. J. Welch (San Francisco: ASP), ASP Conf. Ser., 59, 405--12,  1994.

%\bibitem{ishiguro98}
8.  Ishiguro, M.,  Japanese Large Millimeter and Submillimeter Array, in {\it Advanced technology MMW, radio, and terahertz telescopes}, ed. T. G. Phillips (Bellingham: SPIE), Proc. SPIE, 3357, 244--53, 1998.

%\bibitem{guilloteau98}
 9. Guilloteau, S.,  The LSA Project, in {\it Advanced technology MMW, radio, and terahertz telescopes}, ed. T. G. Phillips (Bellingham: SPIE), Proc. SPIE, 3357, 238--43, 1998.

%\bibitem{MMA Memo Series}
%  The MMA Memorandum Series can be found at http://www.alma.nrao.edu/memos/index.html.
  
%\end{thebibliography}
\cleardoublepage

\end{small}

\end{document}